\documentclass[a4paper,11pt]{article}
\pdfoutput=1 

\usepackage{jcappub} 
\usepackage{natbib}
\usepackage{amsmath}
\usepackage[T1]{fontenc} 
\DeclareMathOperator*{\argmax}{argmax}
\newcommand{\rthis}[1]{\textcolor{black}{#1}}

\title{\boldmath Looking for ancillary signals around GW150914}

\author[a]{Rahul Maroju,}
\author[a]{Sristi Ram Dyuthi,}
\author[a]{Anumandla Sukrutha}
\author[b,1]{and Shantanu Desai, \note{Corresponding author.}}

\affiliation[a]{Department of Electrical Engineering, IIT Hyderabad, Kandi, Telangana-502285, India}
\affiliation[b]{Department of Physics, IIT Hyderabad, Kandi, Telangana-502285, India}

\emailAdd{ee15btech11018@iith.ac.in}
\emailAdd{ee14btech11031@iith.ac.in}
\emailAdd{ee14btech11002@iith.ac.in}
\emailAdd{shantanud@iith.ac.in}

\abstract{We replicated the procedure  in Liu and Jackson~\cite{Liu}, who had found evidence for a  low amplitude signal in the vicinity of  GW150914. This was based upon the large  correlation between the time integral of the Pearson cross-correlation coefficient in the off-source region of GW150914, and the Pearson cross-correlation in a narrow window around GW150914, for  the same time lag between the two LIGO detectors as the gravitational wave signal. Our results mostly agree with those in Liu and Jackson~\cite{Liu}. \rthis{We find the statistical significance of the observed cross-correlation  to be about 2.5 $\sigma$.}
We also used the cross-correlation method to search for short duration signals at all other physical values of the time lag, within this 4096 second time interval, but do not find  evidence for any statistically significant events in the off-source region.}

\begin{document}
\maketitle
\flushbottom
\section{Introduction}
\label{sec:intro}

Three and half  years ago, there was a watershed moment in astronomy with the detection of gravitational waves, by the LIGO detectors, starting with the first binary  black hole merger, viz. GW150914~\cite{GW150914}. Since then, the LIGO-VIRGO collaboration (LVC) has detected a total of seven binary black hole mergers~\citep{LIGO2,LIGO3,LIGO4,LIGO5,LIGO2018} and one binary neutron star merger~\cite{LIGO6}. These detections (which led to the 2017 Nobel Prize in Physics) have provided a plethora of information about astrophysics, nuclear physics, modified theories of gravity, $r$-process nucleosynthesis,  Hubble constant, etc (eg.~\cite{Miller,Kahya,Boran,Moffat17,Miguel,Carson,Metzger} and references therein). Along with the discoveries, the LIGO open science center (LOSC) has publicly initially released the strain data around each of the detections, and later the full science-mode data from the first two observing runs~\cite{O1}.

Nevertheless, a few groups have still raised a few  concerns about some of the  LIGO detections~\cite{Chinese,Creswell}.
For example, Creswell et al~\cite{Creswell} used data from  LOSC around GW150914 to point out that the calibration lines and also the residual noise show a peak in the cross-correlation at the  same time lag as the original signal. However, at least two groups have been unable to  reproduce these results, and therefore dispute these claims~\cite{Moffat,Brown} (see however~\cite{Jackson19}). Conversely, some other groups have found evidence for additional signals and anomalies potentially pointing to new Physics~\cite{Liu,Abedi,Afshordi,Torres}. In Refs.~\citep{Abedi,Afshordi}, the authors have argued for a tentative evidence for echo signals post the merger of the compact objects in both the binary black hole and neutron star events. This work has also received a lot of attention and many groups have tried to reproduce these claims~\cite{Ashton,Brown,Nitz}. In Ref.~\cite{Torres}, the authors have applied a regularization-based denoising technique to the data around GW150914 and found that the reconstructed waveforms disagree at the high frequency end with the GR-based binary black hole template waveforms. Liu and Jackson~\cite{Liu} (LJ16 hereafter), calculated  the Pearson  cross-correlation coefficient for data from Hanford and Livingston detectors in the strain data covering 4096 seconds around  GW150914. Using this analysis, they have found a tentative evidence (3.2$\sigma$) for a signal 10 minutes before GW150914, lasting for about 45 minutes with the same time lag as GW150914.

 Within the LVC, the GW150914 signal was detected by two independent  pipelines: one of which involves a matched filtering search from coalescing binaries and the second search involving a model-independent search for short duration signals of any waveform morphology, known as ``bursts''. Different burst search algorithms, which look for waveform-agnostic transients in LIGO data do not always detect the same triggers~\cite{Stuver}. Therefore, given the importance of gravitational wave detections for physics and astronomy, it is important to verify any concerns about any of the detected events, as well to corroborate any additional signals found by people outside  LVC, using independent analysis techniques. 
 
 Here, we try to reproduce the claims in LJ16, by trying to mimic their procedure as closely as possible. We also look for short-duration transient signals at all time lags using a simple extension of the method used in LJ16. We note that most recently Nitz et al~\cite{Nitz2} have carried out a matched filter search (with coalescing binary signals as templates) over the full O1 data and do not find any additional GW signals, beyond those reported. These signals were also not found in the recent re-analysis by the LVC ~\citep{LIGO2018}.
 While this work was in preparation, Nielsen et al.~\cite{Brown} have written a rebuttal countering the claims in Ref.~\cite{Creswell} (which is follow-up paper to LJ16), where-in they have also used the Pearson cross-correlation coefficient to examine data around $\pm 512$ seconds of data around GW150914 signal.  However, none of these works have tried to reproduce all the claims in LJ16, or replicated their procedure, therein, which is our goal. 

The outline of the paper is as follows. We first briefly recap the results of LJ16 in Sect.~\ref{sec:lj16}. We describe the two independent  data conditioning steps carried out in Sect.~\ref{sec:dc}. We describe the results from these two steps in Sects.~\ref{sec:notch} and \ref{sec:whiten} respectively. \rthis{We then re-analyze the cross-correlations after removing the best-fit GW signal and these results can be found in Sect.~\ref{sec:resid}. Finally, our estimation of significance using Monte-carlo simulations is discussed in Sect.~\ref{sec:sig}.} We conclude in Sect.~\ref{sec:conclusions}.
We have also uploaded our analysis code on {\tt github} and a  web link can be found in Sect.~\ref{sec:conclusions}.

\section{Recap of LJ16}
\label{sec:lj16}
LJ16 downloaded $\pm 2048$ seconds of Hanford and Livingston data (sampled at 16 kHz) around GW150914, starting at GPS time 1126257414. Very simple data conditioning was implemented~\cite{Naselsky} along the same lines as done by the LVC. The data was band-pass filtered between 50 and 350 Hz using a butterworth filter for the primary analysis. However later, they explored the sensitivity of their significance to different bandwidths. Narrow lines (corresponding to power, calibration and violin mode resonances) were removed in the Fourier domain by clipping the amplitudes to zero power~\citep{Creswell}. The final time-series used to search for gravitational waves was then inverse Fourier transformed.  A total of 30 seconds of data was excluded at both the ends to avoid boundary artifacts.
LJ16 then calculated  the function $C(t,\tau,w)$ between the time streams of the two LIGO detectors based upon the Pearson cross-correlation coefficient as a function of the time lag  ($\tau$), duration of the signal ($w$).   It is defined as follows~\cite{Liu}:
\begin{eqnarray}
C(t,\tau,w) &=& Corr(H_{t+\tau}^{t+w+\tau},L_{t}^{t+w}) \\
Corr(x,y) &=&  \frac{\sum_{i} (x_i-\bar{x})(y_i-\bar{y})}{\sqrt{\sum_{i}(x_i-\bar{x})^2\sum_{i}(y_i-\bar{y})^2}}
\end{eqnarray}
\noindent where $x_i$ and $y_i$ are the time-samples over which Pearson cross-correlation coefficient is calculated;  $H_{t+\tau}^{t+w+\tau}$
is the Hanford time-series data from ($t+\tau \rightarrow t+\tau+w$) seconds; and  $L_{t}^{t+w}$ is the same for Livingston data from  ($t \rightarrow t+w$) seconds. We then define $C(t,\tau,w)$ for GW150914, which we refer to as $C_{GW}(\tau,w)$ as follows
\begin{equation}
C_{GW}(\tau,w) = Corr(H_{t_{GW}+\tau}^{t_{GW}+w+\tau},L_{t_{GW}}^{t_{GW}+w}) 
\label{eq:cgw}
\end{equation}
We note that a similar statistic was previously used as a signal consistency test in the post-processing pipeline  for LIGO burst searches during science runs  S2, S3, and  S4~\cite{s2burst,s3burst,s4burst}, under the moniker {\tt r-statistic}~\cite{rstatistic} or {\tt CorrPower}~\cite{corrpower}. 
This pairwise cross-correlation posits similarity of the signals (modulo any offsets in calibration). This {\it ansatz} is valid for the LIGO detectors, as the 4 km Livingston interferometer is maximally aligned with the Hanford detectors  to the extent allowed by Earth's curvature (modulo a sign flip in one of the detectors, which makes the signals anti-correlated)~\cite{LIGONIM}. For mis-aligned detectors, Pearson cross-correlation cannot be used and needs to be replaced by Rakhmanov-Klimenko cross-correlation or similar variants~\citep{Rakhmanov}.

In LJ16, cross-correlation has been run in extremely short-duration  non-overlapping time windows (0.1 seconds) as well as over data stretches as long as 30 minutes. In Ref.~\cite{Brown}, the window duration for calculating cross-correlation was chosen to be 0.2 seconds.
To look for weak-amplitude signals, LJ16  introduced the functional $D(\tau)$, related to the integral of the cross-correlation  and is defined as~\cite{Liu}
\begin{equation}
D(\tau) = \frac{1}{t_2-t_1}\int_{t_1}^{t_2} C(t,\tau,w) dt 
\end{equation}
where $w$ is the duration of the signal. In LJ16, $w$ was chosen to be 0.1 seconds, but we consider $w$ values of both 0.1 and 0.2 seconds. Note that GW150914 lasted for about 0.2 seconds with frequency sweeping from 35  to  150 Hz over this duration.
The main premise in positing the above statistic is that with an increasing window duration, the contribution of noise should decrease, whereas even a weak signal would add up coherently.  They also considered the cross-correlation of $C_{GW}(\tau,w)$ (from Eq.~\ref{eq:cgw}) and 
$D(\tau)$, (where $D(\tau)$ is calculated  over a time interval from $t_1$ to $t_2$), which they refer to as $E(t_1,t_2)$ and is defined as
\begin{equation}
E(t_1, t_2) = Corr\{C_{GW}(\tau,w),D(\tau)\} \:\mathrm{for}\: \tau \in (\tau_1,\tau_2)
\label{eq:E}
\end{equation}
When $D(\tau)$  and $C(t,\tau,w)$ were applied to a time stretch of data, 30 minutes before and after the signal (after excluding $\pm 60$ seconds of data around GW150914), $\argmax \limits_{\tau}  |C(t,\tau,w)| = \argmax \limits_{\tau}   = |D(\tau)| = 6.9$ ms. 
A more precise estimate of the start and stop time of this ancillary  signal (in the GW150914 off-source  interval) was then obtained by adjusting the start and stop time, as well as by searching in narrow-band filters, until $E(t_1,t_2)$  is maximized. The start time of the signal was thereby estimated to be 1280 seconds after the beginning of the starting time stamp (or  about 10 minutes before GW150914) and ends 45 minutes later. The statistical significance was estimated by calculating the above statistics for  20,000 different  time-shifts with absolute value $>$ 10 ms. This estimated statistical significance was  about $3.2\sigma$.

\section{Data Conditioning of GW150914}
\label{sec:dc}
For our analysis, we downloaded 4096 seconds of data around GW150914, sampled at 4096 Hz, using version 2 calibration (dubbed {\tt v2} and released in October 2016), starting at GPS time 1126257414 (Mon Sep 14 09:16:37 GMT 2015). This start time is about 2048 seconds before the GW150914  signal. This  same calibration strain data file was also used in the recent re-analysis in Ref.~\cite{Brown}. Although, the results in LJ16 were obtained using {\tt v1} of the calibration, Ref.~\cite{Brown} has shown that the results are not sensitive to the calibration version. Therefore, we stick to 
{\tt v2} of the calibration data for this analysis.

Data conditioning for the event is done in the same way as described in the original GW150914 tutorial on the LOSC webpage \url{https://www.gw-openscience.org/s/events/GW150914/GW150914_tutorial.html}. Note that on this tutorial webpage, it has been shown the signal is visible using two independent  data conditioning procedures. The first procedure involves whitening in the frequency domain followed by  band pass filtering (in the time-domain using a butterworth filter). Whitening in the frequency domain is mathematically equivalent to dividing the input data in the Fourier domain by its amplitude spectral density. This whitened frequency domain data  is then transformed back to the time domain, which is then used for GW analysis. 
Alternately, the signal is also visible from the same band-pass butterworth filter, followed by application of IIR notch filtering in the time domain to get rid of various lines. As mentioned earlier, this signal was seen by both the burst and compact binary coalescence-based matched filtering search pipelines, which look for inspiral and merger of compact stellar remnants, also known as ``CBC'' within the LVC~\cite{LVCburst,LVCcbc}. The burst search was done using two pipelines, {\tt coherent Waveburst} (cWB)~\cite{Klimenko} and also {\tt omicron-LALInference-Bursts}~\cite{Lynch}.
The {\tt cWB} algorithm uses a linear predictive error filter, which is equivalent to whitened time series~\cite{Klimenko}. The CBC search was also carried out using two pipelines, called {\tt PyCBC}~\cite{Canton} and {\tt GstLAL}~\cite{Cannon}. Both these pipelines employ a whitening filter  in the  data conditioning steps.

As discussed in Sect.~\ref{sec:lj16}, LJ16 have applied a band-pass window followed by  notch windows, obtained using clipping in the frequency domain. We also note that in the {\tt Block-Normal} burst  search pipeline~\citep{Mcnabb}, which was used until S5~\cite{s5glitch}, data conditioning consisted of all three steps (band-passing, whitening and line removal)~\cite{summerscalespsu}.  In some other pipelines, the data has been whitened using a running median~\cite{Hayama}.

Therefore, there is no unique data conditioning procedure followed by the different 
GW search algorithms. Here,  we present our results using both band-pass and notch filtering (for line removal), as well as  using band-pass and whitening. Ref.~\citep{Brown} have also done their analysis using   notch filtering as well as  whitening.
The code for implementing these steps has been described on the aforementioned LOSC webpage. We now briefly describe both these steps.

 We first apply FFT to the full 4096 seconds of data. We then apply a bandpass filter between 50 Hz and 350 Hz to suppress the low-frequency noise (mainly due to seismic activity) and high frequency noise (mainly due to quantum noise)~\cite{advancedligo}. The noise floor shows a steep rise in this frequency range~\cite{ligonoise}, and hence the signal is always band-passed in this intermediate frequency range to search for astrophysical signals.  The band-passing is implemented using a linear  butterworth filter. The LIGO noise spectrum also contains a number of lines in the frequency domain due to mechanical resonances, power line harmonics, etc~\cite{advancedligo}. These have been removed using IIR based time-domain notch filtering, following the algorithm on the above webpage. The list of lines for which the notch filter has been applied include 14, 34.7, 35.3, 36.7, 37.3, 40.95, 60.0, 120.0, 179.99, 304.99, 331.49, 510.02, 1009.99 Hz respectively. The same lines have also been   notch-filtered in a follow-up paper by Creswell et al.~\citep{Creswell}.
 Results from this data conditioning procedure are described in Sect.~\ref{sec:notch}.
 
 For our second set of results, instead of notch-filtering, we applied a whitening filter, by dividing by the noise spectrum in the frequency domain using the same code  as mentioned on the above LOSC tutorial website.
 
 \section{Results from band-passing and notch filtering}
 \label{sec:notch}
Here, we describe our results after the application of band-pass filtering and notching. The corresponding results after the application of the whitening filter, instead of the notch filter shall  be described in Sect.~\ref{sec:whiten}.

\begin{figure}[tbp]
\centering
\includegraphics[width=.5\textwidth]{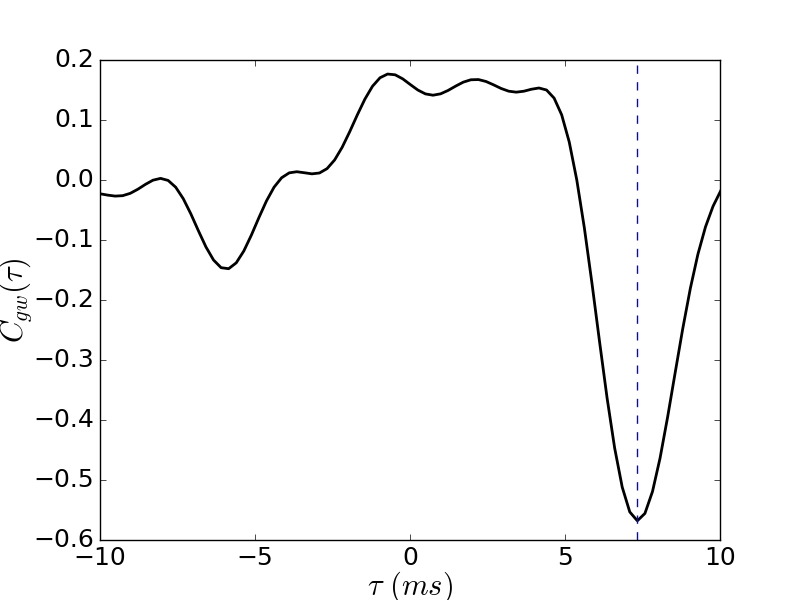}
\caption{\label{fig:1} Cross-correlation  coefficient in a $\pm0.2$ second window around  GW150914 as a function of time-lag between the Hanford and Livingston detectors. The dashed vertical line is for a time lag of 7.33 milliseconds. We note that the data has been downsampled to 4096 Hz, so the minimum time resolution is 0.246 milliseconds. For this plot, the data has been band-passed from 50 to 350 Hz and then notch filtered.}
\end{figure}

\begin{figure}[h]
\centering 
\includegraphics[width=.45\textwidth]{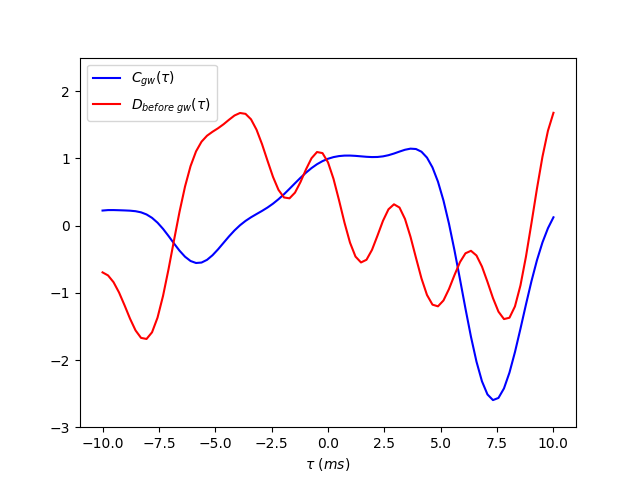}
\hfill
\includegraphics[width=.45\textwidth]{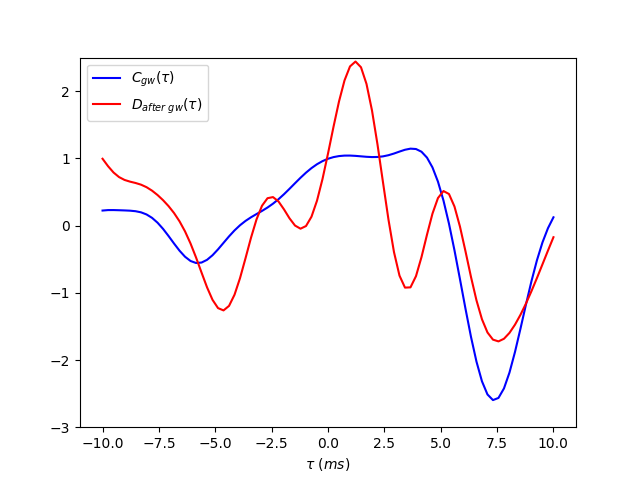}
\caption{\label{fig:2} Value of $C_{GW}(\tau)$ and $D(\tau)$ in a 30 minute time window (starting from GPS time 1126257444) until 60 seconds  before  GW150914 (left panel). The right panel shows the same for a 30 minute time window starting from 60 seconds after GW150914. $D$ has been calculated by integrating over the values of $C(t,\tau,w)$, for $w= 0.2$ seconds. The data conditioning is  the same as in Fig.~\ref{fig:1}. Note however that $C_{GW}$ has been calculated in a $\pm 0.05$ second window around GW150914 and is slightly different compared  to Fig.~\ref{fig:1}.  Both $C_{GW}$ and $D$ have been rescaled so that they have a mean value of zero and rms value of one. We notice that in the right panel  $C_{GW}$ and $D$ are correlated around the same time lag as GW150914 (in agreement with LJ16) but also for time lags close to -10 ms.}
\end{figure}

\begin{figure}[h]
\centering 
\includegraphics[width=.45\textwidth]{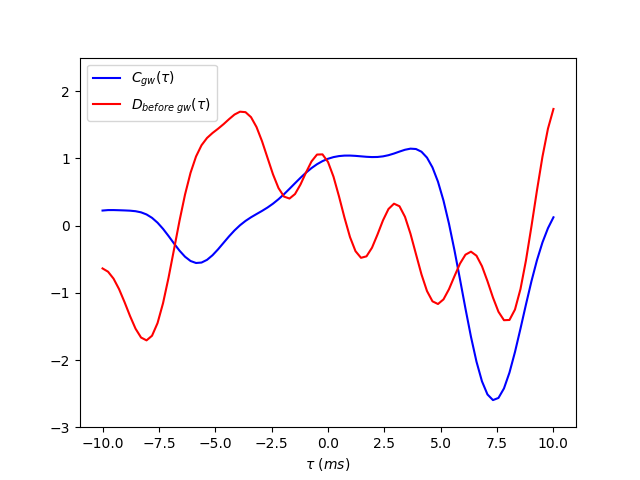}
\hfill
\includegraphics[width=.45\textwidth]{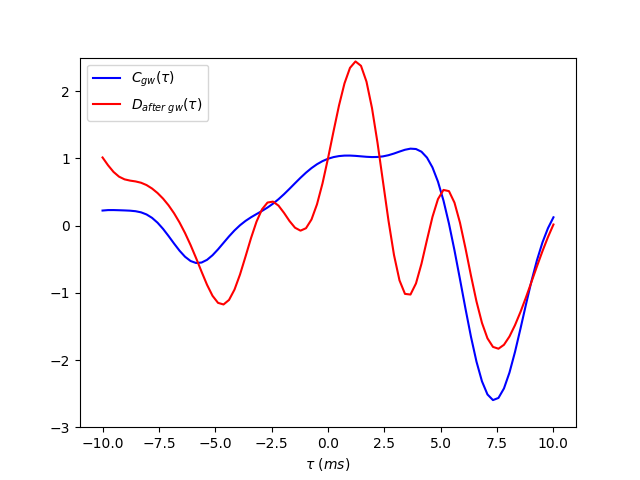}
\caption{\label{fig:3} Same as Fig.~\ref{fig:2}, except that $D$ has been calculated for a 0.1 second window. Our conclusions are  same as in Fig.~\ref{fig:2}.}
\end{figure}

Similar to LJ16, we exclude the first and last 30 seconds of the data. Using this data, we first calculate  $C_{GW}(\tau,w)$ as a function of the time lag ($\tau$) between the two LIGO detectors in a  $\pm$ 0.2 second window around   GW150914 corresponding to the GPS time of  1126259462.42 (Sep 14 09:50:45). Since we have downsampled the data to 4~kHz, the minimum resolution in the time lags we can have is 0.246 milliseconds. 
This plot can be found in Fig.~\ref{fig:1} and is similar to Fig.~1 of LJ16 or the corresponding plot in Fig.~\cite{Brown}. One can see a sharp dip for a time lag close to   7.33 ms corresponding to the signal. Note that unlike LJ16, we do not find a minimum value of $\tau$ closer to 6.9 ms. The peak absolute value of $C$ is at $\sim 7.33$ ms, closer to the value found in Ref.~\citep{Brown}. The uncertainty in the peak time is about 0.246 milliseconds, which is limited  by the  downsampled frequency of 4~kHz.
Similar to LJ16, we then calculated  $D(\tau)$ (from $C(t,\tau,w)$) for a long time stretch of data spanning 30 minutes  before and after the  GW signal, after excluding the 
data within $\pm 60$ seconds around GW150914. We then overlay this along with $C_{GW}$ for $w=\pm 0.05$ seconds around GW150914. Note that for this purpose similar to LJ16, both $C_{GW}$ and $D$ have been rescaled, so that  they have a mean value of zero and a rms value of one. To calculate $D$, one needs to integrate $C(t,\tau,w)$ over a window duration. For the calculation of $D$, we calculated $C$ in  both 0.1 and 0.2 second windows.
These plots for  $C_{GW}$ and $D$  using both the windows can be found in Figs.~\ref{fig:2} and \ref{fig:3}. 
We find roughly the same behaviour as in Fig.~1 of LJ16. In the panels showing the data after GW150914, we find a near-maximal correlation (similar to LJ16) between 
$C_{GW}$ and $D$ at the same time lag as GW150914. However, we also find a similar correlation at the opposite end for $\tau$ between -7 and -10~ms. Therefore, it is not immediately obvious if the strong correlation between $C_{GW}$ and $D$ at the same time lag as GW150914 is pointing to a post-merger signal.

\begin{figure}[tbp]
\centering
\includegraphics[width=.5\textwidth]{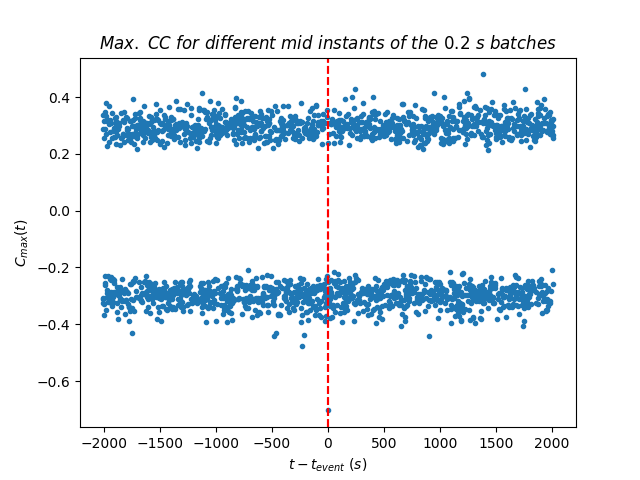}
\caption{\label{fig:4} Maximum values of the cross-correlation  coefficient ($C$) after scanning over  time lags from -10 to 10 milli-seconds  in each of these windows,  over a time stretch of about 4000 seconds encompassing the  GW150914 signal ($t-t_{event}=0$ in the above 
plot). Data conditioning is same as in Fig.~\ref{fig:1}. We find that the maximum absolute value of $C$ occurs at the peak of the GW150914 signal and there is no other statistically significant peak outside the vicinity of GW150914. The maximum (absolute) value of cross correlation is -0.703 and occurs in the batch with mid-instant 1126259462.48 with  $\tau$ equal to  7.32 ms, which corresponds to GW150914. In the off-source region (ignoring 60 seconds around the GW event), the largest (absolute) value of $C$ occurs 233 seconds before GW150914 with a value of -0.47.}
\end{figure}

As an extension of this technique, we then look for additional short duration signals lasting 0.1 and 0.2 seconds  using the cross-correlation method, which may appear at arbitrary time lags. Furthermore, if there is a persistent long-duration signal, then $C$ should have the same  value across many 0.1/0.2 second chunks. 
For this purpose, we calculated the cross-correlation in each 0.1/0.2 second chunk for the entire 4096 seconds of LIGO data, for time lags ranging from -10 to 10 milliseconds. These plots can be found in Fig.~\ref{fig:3} and Fig.~\ref{fig:4}.
The maximum (absolute) value of $C$  is seen at the peak of GW150914. We do not find any values of $C$ as large as that for GW150914  at any other times.  For the 0.1 second window, the maximum cross correlation in the off-source region has a value  of 0.63 at the GPS time of  1126257962.4, about 548 seconds from the start of the time series or 1500 seconds before GW150914.
For the 0.2 second window, the maximum absolute cross correlation  in the off-source region equal to  -0.47 occurs at the GPS time of 1126259229.4, about 1815 seconds from the start of our time series, or 233 seconds before GW150914. The maximum  (absolute) values of $C$  at the peak of GW150914 are -0.7 and -0.87 for the 0.2 and 0.1 second windows respectively.

Therefore, with this data conditioning procedure, we roughly agree with the results in LJ16, although we also find a strong correlation between $C_{GW}$ and $D$ for time lags close to -10 ms, which is far from the time lag seen for GW150914. However, when we scanned the full 4096 seconds of data around GW150914, we do not find any other value of $C$
calculated in small duration windows, whose absolute value is as large as GW150914, for both the windows.

\subsection{Correlation between $C_{GW}$ and $D$}
\label{sec:E}
To further pin down the exact start and end time  of the signal, LJ16 calculated  $E$ (cf.~\ref{eq:E}) by calculating the cross-correlation between $C_{GW}$ and $D$. They find that $E$ is maximized, when the signal starts at 1280 seconds and ends at 4050 seconds from the start of the downloaded data segment. We calculated the values of $E$ for the same starting and ending times as LJ16 from 1280 to 4050 seconds.  Note however that unlike LJ16, we have excluded the data within $\pm 60$ seconds  around GW150914, for evaluating $D$, in order to avoid any bias from the GW signal in the final value for $E$\footnote{LJ16 found  that excluding the data around GW150914 for the purpose of calculating $E$, does not change the results in Section III B if their paper (H. Liu, private communication).}. Therefore, instead of  $E(1280,2048)$, $E(2048,4050)$, and $E(1280,4050)$, we evaluated
$E(1280,1988)$, $E(2108,4050)$, and $E(1280,1988) \bigcup E(2108,4056)$. 
Given the similarity in the trends of $C_{GW}$ and $D$ for time lags close to -10 ms, we also checked the value of $E$ for $\tau \in (-10,-6)$ ms \rthis{and $\tau \in (-9,-5)$ ms}. These values  of $E$ can be found in Table~\ref{intro_table}.
Similar to LJ16, we find that $E(2108,4050)$ is largest for $\tau \in$  (5,9) ms,  and  for $\tau \in$  (-10,10) ms, $E$ is largest for $t_1=1280$ and $t_2=4050$, even after excluding $\pm$ 60 seconds around GW150914. Therefore, our results for $E$ are broadly in agreement with those in LJ16, even after excluding $\pm 60$ seconds of data around  GW150914 signal. However, we also 
find a value for $E(2108,4050)$  very close to one for $\tau \in (-10,-6)$ ms. If there was an external ancillary post-merger signal at the same time lag as GW150914, then we cannot think of any apriori reason why $E$ should be close to one for a time lag close to -10 ms.

 We then repeat the test carried out in LJ16 by band-passing the data into contiguous 20 Hz bands between 30 Hz and 150 Hz. These values for $E$ in different bands  can be found in Table~\ref{table:2}. Our results agree with Table II in LJ16.  We also find that the correlations in 50-70 Hz and 90-110 Hz are very close to 1.
 
 Therefore,  to summarize, our results for $E$ broadly agree with those in LJ16, for  the same time lag as GW150914 (7.5 ms). But we also find a large value for $E$ at time lags between $-10$ and $-6$ ms.

\begin{figure}[tbp]
\centering
\includegraphics[width=.5\textwidth]{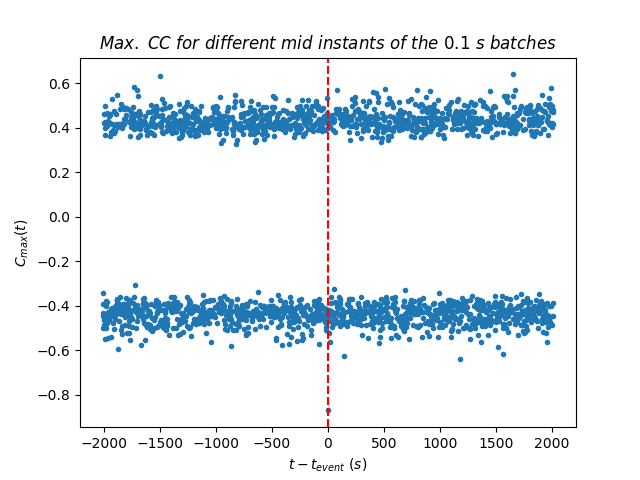}
\caption{\label{fig:5} Same as Fig.~\ref{fig:4}, except that $C$ was calculated in 0.1 second window. Similar to Fig.~\ref{fig:3}, the peak occurs during the occurrence of GW150914 and no other value of $C$ is close. The maximum (absolute) cross correlation is equal to -0.88 and occurs in the batch with mid-instant 1126259462.4 with a corresponding $\tau$ of 7.32 ms,  which corresponds to GW150914. In the off-source region, the next largest (absolute) value of $C$ occurs 1500 seconds before GW150914 with a value of 0.63.}
\end{figure}

\begin{table}
\renewcommand{\arraystretch}{1.3}
\centering
\begin{tabular}{|c|c|c|c|}
\hline   {} & {\bf E(1280, 1988)} & {\bf E(2108, 4050)} & {\bf E(1280,1988)$\bigcup$(2108, 4050)}   \\ \hline
\textbf{\boldmath$\tau \in $  (5 ms, 9 ms)} & 0.147 & 0.946 & 0.831 \\ \hline
\textbf{\boldmath$\tau \in $ (-10 ms, -6 ms)} & -0.976 & 0.969 & -0.523
\\ \hline 
\textbf{\boldmath$\tau \in $ (-9 ms, -5 ms)} & -0.989 & 0.898 & -0.259
\\ \hline 
\textbf{\boldmath$\tau \in $ (-10 ms, -10 ms)} & 0.371 & 0.664 & 0.775
\\ \hline 

\end{tabular} 
\caption{$E$ (defined in Eq.~\ref{eq:E}) over different time windows (columns) and for different time lags (rows). This is analogous to Table I of LJ16~\cite{Liu}, except that we have removed data within $\pm$ 60 seconds of GW150914 for calculating $D$. Our results mostly agree with the corresponding table in LJ16.}
\label{intro_table}
\end{table}

\begin{table}
\renewcommand{\arraystretch}{1.3}
\centering
\begin{tabular}{|c|c|}
\hline \bf{Bandpass range (Hz)} & \bf{E(-10 ms, 10 ms)} \\ \hline
30-50 & 0.813\\ \hline
50-70 & 0.913\\ \hline
70-90 & 0.212\\ \hline
90-110 & 0.996\\ \hline
110-130 & 0.426\\ \hline
130-150 & 0.514\\ \hline
\end{tabular} 
\caption{$E$(-10 ms, 10 ms) in various sub-bands for the time interval (1280, 4050) excluding $\pm$ 60 seconds around the GW event. This is analogous to Table III of LJ16~\cite{Liu} and our results are mostly in agreement.}
\label{table:2}
\end{table}





\section{Results after band-pass filtering and whitening}
\label{sec:whiten}

\begin{figure}[tbp]
\centering
\includegraphics[width=.5\textwidth]{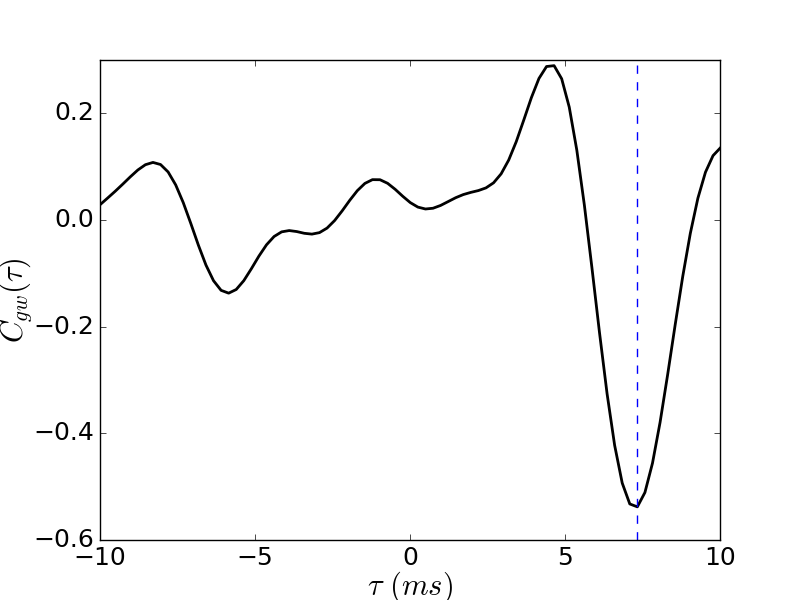}
\caption{\label{fig:1whiten} Same as in Fig.~\ref{fig:1}, except instead of notch filtering, data has been whitened. Again, we  find a  (negative excursion) peak at 7.33 milliseconds at the same time as the GW signal.}
\end{figure}

\begin{figure}[h]
\centering
\includegraphics[width=.45\textwidth]{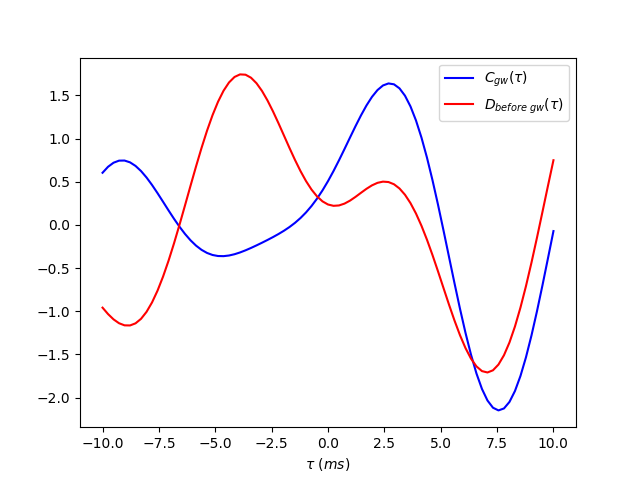}
\hfill
\includegraphics[width=.45\textwidth]{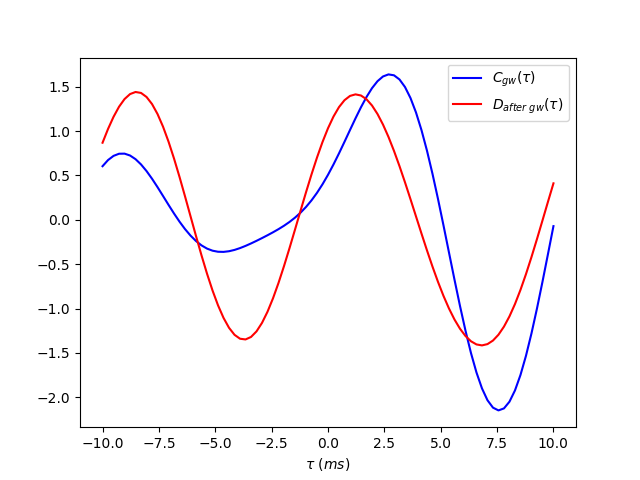}
\caption{\label{fig:2whiten} Same as Fig.~\ref{fig:2}, except that the data has been whitened instead of notch-filtered. $D$ has been calculated in 0.1 second batches. The  correlation between $C_{GW}$ and $D$ trends for the figure in the right panel  is  same as in Fig.~\ref{fig:2}.}
\end{figure}

\begin{figure}[h]
\centering
\includegraphics[width=.45\textwidth]{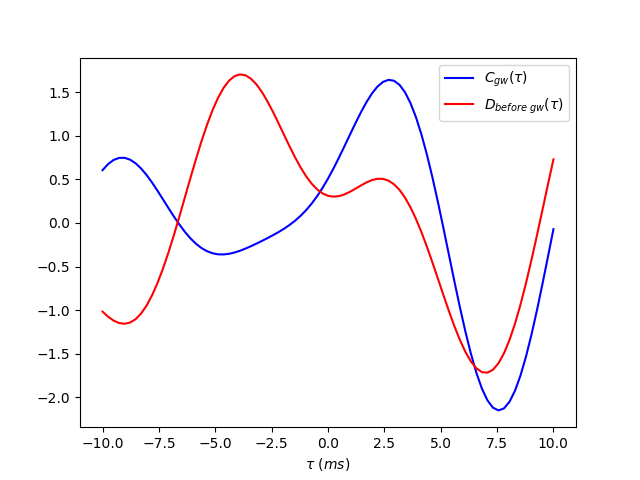}
\hfill
\includegraphics[width=.45\textwidth]{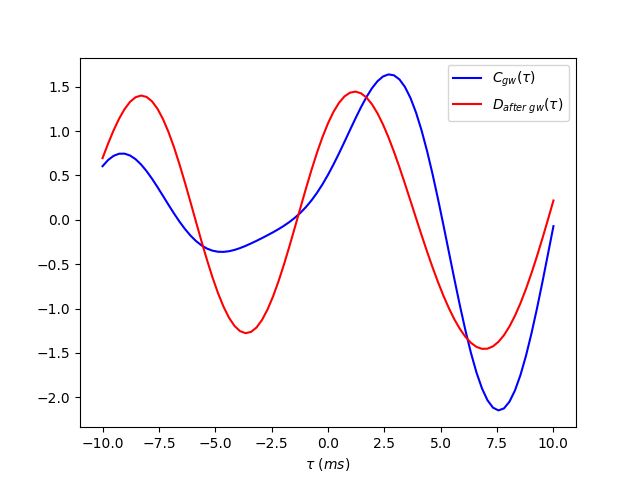}
\caption{\label{fig:2whitenpointtwo} Same as Fig.~\ref{fig:2whiten}, except that $D$ has been calculated in 0.2 second windows. Our conclusions are same as in Fig.~\ref{fig:2whiten}.}
\end{figure}

We now go through  the same  exercise  as in Sect.~\ref{sec:notch}, after skipping the  notch filtering step, but replacing it by whitening the data. We first calculate the cross-correlation in a $\pm0.2$ second window around GW150914 , and the resulting figure can be found in Fig.~\ref{fig:1whiten}. Again, we see a peak at around 7.33 milliseconds. We then calculate $C_{GW}$ and $D$ for the same time window as in Figs.~\ref{fig:2} and \ref{fig:3}. These plots for 0.1 and 0.2 second signal durations  can be  found in Figs.~\ref{fig:2whiten} and ~\ref{fig:2whitenpointtwo}. Again, the qualitative trends are similar to those in Figs.~\ref{fig:2} and ~\ref{fig:3}, and similar to LJ16, we see a strong correlation between $C_{GW}$ and $D$  at the same time lag as GW150914, for  a 30 minute stretch of  data after GW150914. However, for the same data, we also notice a  correlation between $C_{GW}$ and $D$
for $\tau$ between -7.5 and -10.0 ms.

\begin{figure}[tbp]
\centering
\includegraphics[width=.5\textwidth]{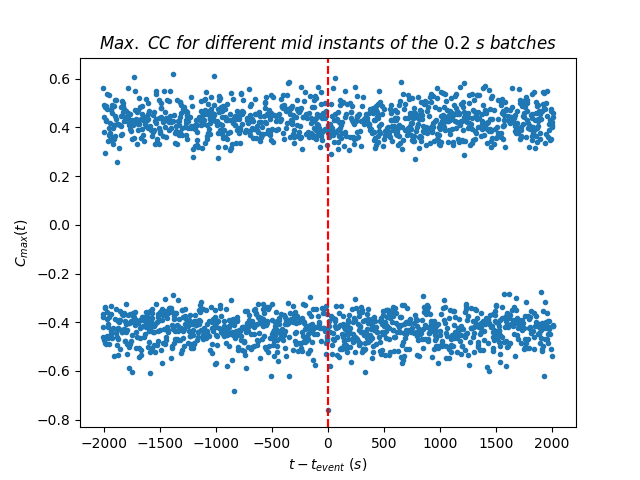}
\caption{\label{fig:9} Same as Fig.~\ref{fig:4}, except that the data has been whitened, instead of applying a notch filter. The largest (absolute) value of $C$ is for  GW150914. This maximum (absolute) cross correlation is equal to -0.76 and occurs in the batch with mid-instant GPS time of 1126259462.5 seconds, with a corresponding $\tau$ of 7.57 ms. In the off-source region, the largest (absolute) value of $C$ occurs 839 seconds before GW150914 with a value of -0.68.
}
\end{figure}

\begin{figure}[tbp]
\centering
\includegraphics[width=.5\textwidth]{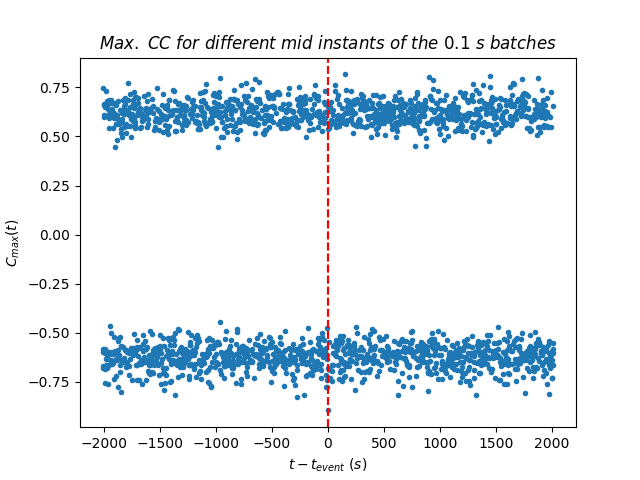}
\caption{\label{fig:10} Same as Fig.~\ref{fig:9}, except that  $C$ was calculated in 0.1 second window. The only peak is seen at the location of GW150914, as in the other plots. The maximum absolute cross correlation is -0.895 and occurs in the batch with mid-instant GPS time of  1126259462.4 seconds for $\tau$ equal to 7.56 ms, which corresponds to the GW150914. In the off-source region, the next largest (absolute) value of $C$ occurs 274 seconds before GW150914 with a value of -0.82.}
\end{figure}

Finally, the values of $C$  in 0.1 and 0.2 second chunks are shown in Figs.~\ref{fig:9} and \ref{fig:10}. We do not find any other time interval with a value of $C$ as large as that of GW150914.
In the off-source region and for a 0.1 second time interval, the  maximum  (absolute) cross correlation is about  -0.82 and occurs at a GPS time of  1126259188.7, which is about 1774 seconds from the start of the time-series, or 274 seconds before GW150914.   For the 0.2  second window, the  maximum (absolute) cross correlation is  about  -0.68 and occurs at the GPS time of  1126258623.6, which is about 1209 seconds from the start of the time series, or 839 seconds before GW150914. The maximum  (absolute) values of $C$  at the peak of GW150914 are -0.76 and -0.82 for the 0.2 and 0.1 second windows respectively. 

Therefore, even after the application of a whitening filter (instead of notch filter) as part of the data conditioning step, the trends in $C$ and $D$ are in agreement with  those in LJ16, as well as our previous results discussed in Sect.~\ref{sec:notch}.  Based upon the values of $C$ in 0.1 and 0.2 second windows, we do not find evidence for any additional short duration signals in the off-source region of GW150914.

\section{Cross-correlations after removing the GW signal}
\label{sec:resid}
\begin{figure}[tbp]
\centering
\includegraphics[width=.5\textwidth]{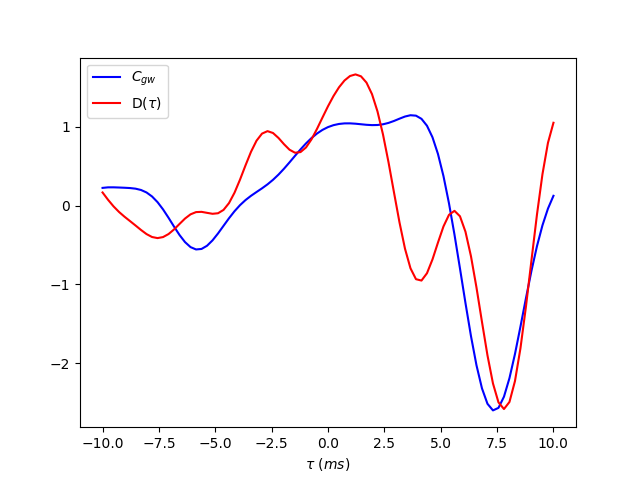}
\caption{\label{fig:CCresid} \rthis{$D(\tau)$ in a time window from 1280 to 4050 seconds,  by including the full data , after   subtracting the signal around GW150914 using the best-fit matched filtering template, so that the residual is consistent with noise. We find that similar to previous such figures, $C_{gw}(\tau)$ is strongly correlated with $D(\tau)$ for the same time lag as GW150914. }} 
\end{figure}

\rthis{In Sections ~\ref{sec:notch} and \ref{sec:whiten}, we have calculated the cross-correlation between $D$ and $C_{gw}$, after removing $\pm$ 60 seconds of data around GW150914  in the calculation of $D$. Here, we redo the cross-correlation after including these 60 seconds, but after subtracting the gravitational wave signal around GW150914. For this purpose, we  subtracted  the matched filter template from the filtered data  in a 0.45 seconds window starting from GPS time 1126259462  (Sep 14 09:50:45 GMT 2015)} \footnote{For this analysis, we directly used the residuals provided by the LOSC, which are available at \url{https://www.gw-openscience.org/s/events/GW150914/P150914/fig1-residual-H.txt} and \url{fig1-residual-L.txt}}, \rthis{so that the full datastream used in the calculation of $D$ is consistent with noise. We show the trends for  $D$  (after using this data with the  GW signal subtracted) in a time window from 1280 to 4050 seconds, along with  $C_{GW}$ in Fig.~\ref{fig:CCresid}. Similar to before, we see a strong correlation between $D(\tau)$
and $C_{GW}(\tau)$.}

\rthis{We then repeat the same exercise as in Sect.~\ref{sec:E} and calculate $E$ in the same time windows using this signal-subtracted data around GW150914. These results can be found in Table~\ref{table:3}. The results are the same as in Table~\ref{intro_table}. We still see a value for $E(1280,4050)$ close to one for
 $\tau \in (5,9)$ ms and $\tau \in (-10,10)$ ms.}

\begin{table}
\renewcommand{\arraystretch}{1.3}
\centering
\begin{tabular}{|c|c|c|c|}

\hline   {} & {\bf E(1280, 2048)} & {\bf E(2048, 4050)} & {\bf E(1280, 4050)}   \\ \hline
\textbf{\boldmath$\tau \in $  (5 ms, 9 ms)} & 0.118 & 0.949 & 0.838 \\ \hline
\textbf{\boldmath$\tau \in $ (-9 ms, -5 ms)} & -0.954 & 0.932 & -0.743
\\ \hline 
\textbf{\boldmath$\tau \in $ (-10 ms, -6 ms)} & -0.940 & 0.992 & 0.113
\\ \hline
\textbf{\boldmath$\tau \in $ (-10 ms, -10 ms)} & 0.320 & 0.652 & 0.755
\\ \hline 

\end{tabular} 
\caption{\rthis{Similar to Table~\ref{intro_table}, except that we have removed the GW signal from the data instead of ignoring the data around the GW150914, for calculating $D$. Our values for $E$ are approximately the  same as in Table~\ref{intro_table}.}}
\label{table:3}
\end{table}

\section{Estimation of significance}
\label{sec:sig}
\rthis{To estimate the  statistical significance for our values of $E$ in Sections~\ref{sec:E} and \ref{sec:resid}, we time-shift the data from one of the detectors using unphysical time lags, in order to generate Monte-Carlo simulations for a pure noise floor. A similar procedure is usually used to estimate the significance of all thw GW detections by the LVC collaboration (cf. Ref.~\cite{GW150914} for GW150914), and was also used by LJ16 to estimate the significance of their estimated values of $E$.}

\begin{figure}[tbp]
\centering
\includegraphics[width=.5\textwidth]{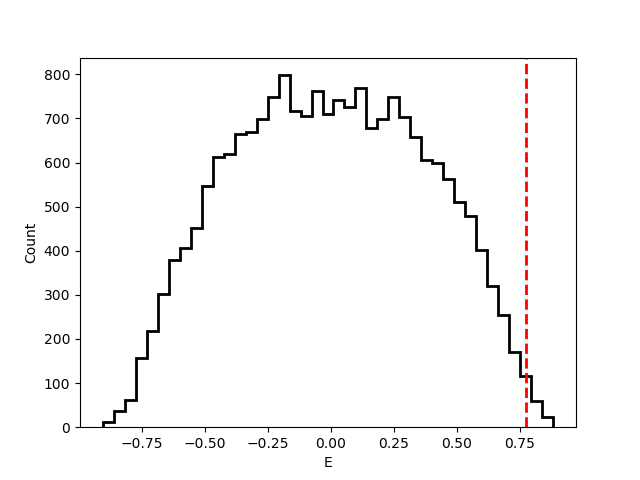}
\caption{\label{fig:Ehist} \rthis{A histogram of $E (1280, 4050)$ calculated
by shifting the Livingston data using unphysical time lags from -600  to -4 seconds in 20,000 steps. For each of these lags, $E$ is calculated for $\tau \in (-10,10)$ ms
The number of realizations  resulting in $E$ exceeding the thresholds 0.775, 0.755, 0.84 are 126, 192 and 22 respectively corresponding to False alarm probabilities of $6.3 \times 10^{-3}$, $9.6 \times 10^{-3}$, and $1.1 \times 10^{-3}$ respectively.}} 
\end{figure}

\rthis{Here, we consider the Hanford data from 1280 to 4050 seconds. We then analyze the Livingston data with the same duration, but time-shifted with respect to Hanford data from ($-600,-4$) seconds in 20,000 steps.\footnote{Note that in LJ16 to estimate the statistical significance, only one second of Hanford/Livingston data separated by unphysical time lags has been used (H. Liu, private communication).} For each value of this time-shifted pair of data, we calculate $D(\tau)$ for $\tau \in (-10,-10)$ ms. Each $D(\tau)$ is cross-correlated with $C_{gw}(\tau)$ to evaluate $E$. Each value of $E$ using this combination of data with unphysical time lags, can be considered as the value for  ``null-stream'' data with no signal and any value close to one, can only be because of a statistical fluctuation. The observed values of $E$  for each of these realizations can be found in Fig.~\ref{fig:Ehist}. } 

\rthis{The number of realizations, which exceed our observed value of $E$ equal to 0.775 (cf. Sect.~\ref{sec:E}) is equal to 126 out of 20,000, corresponding to a  false alarm probability (FAP) of $6.3 \times 10^{-3}$ or a significance of  2.5$\sigma$. The number of realizations for which we get a  value of $E$ greater then 0.755 (which we get after including the subtracting the GW signal  around GW150914, as outlined in Sect.~\ref{sec:resid}) is 192 out of 20,000, for a false alarm probability (FAP) of $9.6 \times 10^{-3}$ or  a significance of 2.34$\sigma$.  Finally, we get a value of $E$, greater than that obtained in LJ16 (viz. 0.84) for 22 out of 20,000 realizations, for a FAP of $1.1 \times 10^{-3}$ corresponding to a significance of 3.06$\sigma$. This significance is about the same as found in LJ16.}

\section{Conclusions}
\label{sec:conclusions}

We replicate the procedure in LJ16~\citep{Liu}, to see if we can independently corroborate their claim of  a long duration low-amplitude  signal at $3.2\sigma$ in the vicinity of GW150914 at the same time lag (as GW150914). We implemented  two different data conditioning procedures
for data within $\pm$ 2048 seconds  of GW150914 (which lasted for about  0.2 seconds from 35 to 150 Hz). The only difference with respect to the analysis in LJ16 is that we have used version 2 of the calibrated strain data and applied notch filtering in the time domain, instead of clipping in the frequency domain as in LJ16. Similar to LJ16,  we calculated the Pearson cross-correlation coefficient ($C(t,\tau,w)$), and also its integral $D(\tau)$ for 0.1 and 0.2 second windows across the full $\pm$ 2048 seconds of data. LJ16 had found that  $D(\tau)$ in this off-source region is maximally correlated with
$C_{GW}(\tau,w)$  for the same time lag as GW150914, where $C_{GW}(\tau,w)$ is the Pearson cross-correlation coefficient for a time-window $w$ around GW150914. They 
showed that this ancillary signal lasted between  1280 and 4050 seconds, where the zero point corresponds to 2048 seconds before GW150914. We roughly find the same trends   in the behaviour of $C_{GW}$ and $D$, using both the data conditioning procedures. Similar to LJ16, we find  a strong correlation in  $C_{GW}$ and  $D$ at the same time lag as GW150914 (around 7.5 ms) in a 30-minute window after GW150914.  \rthis{We then reanalyzed the cross-correlation between $C_{GW}$ and $D$, after including the full data stream, but after removing the best-fit template around the GW signal. We find that the values of $E$ for time lags close to 7.5 ms is about the same as before.  We also the statistical significance of obtaining a value greater than our observed cross-correlation 
to be about $2.5\sigma$, which is roughly comparable to the significance
of $3.1\sigma$ found in LJ16.}

As an extension of this method,  we also evaluated the cross-correlation coefficient in non-overlapping 0.1/0.2 second windows in order to discern the existence of any other  short duration signal in the off-source region, for all physical time lags. We do not find any  other data stretch, which shows a value of $C$  in this region, which could be deemed  statistically significant.

We have made available our Python scripts containing all the codes to reproduce the results at \url{https://github.com/RahulMaroju/GW150914-analysis}


\acknowledgments

This research has made use of data, software and/or web tools obtained from the Gravitational Wave Open Science Center (https://www.gw-openscience.org), a service of LIGO Laboratory, the LIGO Scientific Collaboration and the Virgo Collaboration. LIGO is funded by the U.S. National Science Foundation. Virgo is funded by the French Centre National de Recherche Scientifique (CNRS), the Italian Istituto Nazionale della Fisica Nucleare (INFN) and the Dutch Nikhef, with contributions by Polish and Hungarian institutes. We thank Hao Liu for explaining in detail all the technical details of LJ16. We are grateful to  Soumya Mohanty for the comments on the manuscript and to Malik Rakhmanov for helpful correspondence.



\bibliography{refs}
\end{document}